\documentclass{elsart}

\usepackage{graphicx}
\usepackage{epsfig,array}
\usepackage{booktabs}

\newcommand{\MSbar}{\hbox{$\overline{MS}$\ }}

\newcommand{\beq}{\begin{equation}}
\newcommand{\eeq}{\end{equation}}
\newcommand{\beqa}{\begin{eqnarray}}
\newcommand{\eeqa}{\end{eqnarray}}

\newcommand\tab[1]{{\footnotesize {\bf Table}~[{\bf\ref{#1}}]}}

\newcommand{\sla}[1]%
        {\kern .25em\raise.18ex\hbox{$/$}\kern-.75em #1}
\newcommand{\mybar}[1]%
        {\kern 0.8pt\overline{\kern -0.8pt#1\kern -0.8pt}\kern 0.8pt}

\newcommand\mycaption[1]{\caption{\footnotesize \sf #1}} 

\begin{document} 

\begin{frontmatter}

\title{Heavy quark masses in the continuum limit of quenched Lattice QCD}

\author[romeII]{G. M. de Divitiis}
\author[romeII]{M. Guagnelli}
\author[fermi]{F. Palombi}
\author[romeII]{R. Petronzio}
\author[romeII]{N. Tantalo}

\address[romeII]{University of Rome ``Tor Vergata'' and INFN sez. RomaII, 
	Via della Ricerca Scientifica 1, 
	I-00133 Rome}

\address[fermi]{Enrico Fermi Research Center, 
	Via Panisperna 89a, I-00184 Rome}

\begin{abstract}
We compute charm and bottom quark masses in the \emph{quenched}
approximation and in the continuum limit of lattice QCD. 
We make use of a
step scaling method, previously introduced to deal with two scale
problems, that allows to take the continuum limit of the lattice data. 
We determine the RGI quark masses and make the connection to the
\MSbar scheme. 
The continuum extrapolation gives us a value $m_b^{RGI} = 6.73(16)$ GeV for the $b$--quark 
and $m_c^{RGI} = 1.681(36)$ GeV for 
the $c$--quark, corresponding respectively to $m_b^{\overline{MS}}(m_b^{\overline{MS}}) = 4.33(10)$ GeV
and $m_c^{\overline{MS}}(m_c^{\overline{MS}}) = 1.319(28)$ GeV. 
The latter result, in agreement with current estimates, is for us a check of the method.
Using our results on the heavy quark masses we compute the mass of the $B_c$
meson, $M_{B_c} = 6.46(15)$ GeV.
\end{abstract}


\end{frontmatter}

\section{Introduction}
\label{sec:introduction}

Quark masses are fundamental parameters of the QCD Lagrangian. Their accurate knowledge is required in order to 
give quantitative predictions of fundamental processes. A direct experimental measurement of quark masses is
not possible because of confinement, and their determination can only be inferred from a theoretical 
understanding of the hadron phenomenology. The calculations can be numerically performed through different 
strategies, depending upon the quark flavor and the available computational facilities. Accurate non--perturbative 
measurements of the $u$, $d$, $s$ and $c$ quark masses have been obtained from a straightforward comparison of hadron 
spectroscopy and lattice QCD predictions (see \cite{Wittig:2002ux} for a recent review). 

The situation is different for the $b$ quark, with present
computers capabilities. In principle, the $b$ quark mass could be
extracted, similarly to lighter flavors, by looking 
at the heavy--light and heavy--heavy meson spectrum. 
However, heavy--light mesons are characterized by the presence of two 
different  scales, i.e. $\Lambda_{QCD}$, that sets the wavelengths of
the light quark, and the heavy $b$--quark mass.  
Managing these two scales in a naive way 
would require a very large lattice. 
Indeed, this should contain enough
points to properly resolve the propagation 
of the heavy quark and to make the light quark insensitive to
finite volumes effects.  
A typical size  would be $O(100^4)$ points, hardly affordable in terms of current memory and CPU
time. The heavy--heavy mesons case is simpler, being characterized by
a single scale. 
However, the heaviness of the bound state
makes the exponential decay of the meson correlation functions too
fast and the ground state effective
mass, at large time separations, cannot be disentangled from the numerical noise
(at least on single precision architectures).
 
The task of determining the $b$--quark mass has been faced in literature by resorting to some
approximations of the full theory, 
e.g. HQET on the lattice \cite{Gimenez:2000cj}, 
lattice NRQCD \cite{Gray:2002vk}, or QCD sum rules \cite{Bauer:2002sh,Battaglia:2002tm}. 
A novel approach recently introduced, based on the non--perturbative renormalization of the
static theory and its matching to QCD, has lead to 
a very precise determination of the $b$--mass in the static approximation \cite{Sommer:2002en,Heitger:2003xg}. 

An alternative approach to the bottom quark physics, based on finite size scaling, has been proposed in a previous 
paper \cite{Guagnelli:2002jd}, where it has been applied to the
heavy--light meson decay constants. 
The main advantages of this
{\it step scaling method} (SSM) are that the entire computation is
performed with the relativistic QCD Lagrangian and that
the continuum limit can be taken, avoiding the unfeasible
direct calculation.
In order to implement the SSM, a finite size scheme is required, 
and we adopt the Schr\"odinger Functional (SF) as the most useful
framework. This paper is devoted to apply the  
SSM to the study of the heavy meson spectrum in order to obtain from it
the first determination of the $b$--mass in the 
continuum limit of lattice regularization and quenched approximation. 

The paper is organized as follows: section~\ref{sec:stepscalingmethod} introduces the main ideas underlying the SSM; in section~\ref{sec:framework} details are 
given on its specific implementation through the SF. In section~\ref{sec:hqet} we discuss
the predictions of HQET on the heavy--light step scaling functions. 
Section~\ref{sec:numresults} includes the analysis and numerical results for each step. 
Section~\ref{sec:physresults} contains the results 
on a physical volume and in the continuum limit. The conclusions are drawn in section~\ref{sec:conclutions}. 

\section{The step scaling method}
\label{sec:stepscalingmethod}

The SSM has been designed in order to deal with two scale problems in lattice QCD \cite{Guagnelli:2002jd}, and it has
been shown to work successfully in a first estimate of the $B$--meson decay constant at finite lattice spacing. 
A detailed explanation of the method can also be found in \cite{Guagnelli:2002adven}, where the calculation 
of the $B$--meson decay constant has been
performed in the continuum limit. Here the SSM is reviewed to set the notation and explain why it can be used
also in the case of the $b$--quark mass.

On very general grounds, the method can be conveniently used in order to compute physical observables ${\mathcal O}(E_\ell,E_h)$
depending upon two largely separated energy scales $E_\ell$ and $E_h$ ($E_\ell \ll E_h$), where 
direct simulation, without
introducing big lattice artifacts, would require a very demanding computational effort. The main assumption of the 
method is that the finite size effects affecting $\mathcal O$ have a mild dependence upon 
variations of the high energy scale and are 
controllable from a numerical point of view. 
Finite size effects
can be obtained from 
the ratio, $\sigma_O$, of the observable $\mathcal O$ computed on two different finite
volumes, e.g. $L$ and $2L$, 
\beq
\sigma_{\mathcal O}(E_\ell, E_h, L) = \frac{\left.{\mathcal O}(E_\ell, E_h)\right|_{2L}}{\left.{\mathcal O}(E_\ell, E_h)\right|_{L}}
\eeq
The step scaling function $\sigma_{\mathcal O}$ should simplify in the
region where $E_h\gg E_\ell$. In 
principle, a total decoupling of $E_h$ would determine an absolute insensitivity to variations
of this scale
\beq
\sigma_{\mathcal O}(E_\ell, E_h, L) \simeq \sigma_{\mathcal O}(E_\ell, L), \qquad E_h \gg E_\ell
\label{eq:scaling} 
\eeq
In practice, $E_h$ never completely decouples, but the mild residual
dependence can be suitably parametrized.
A typical situation is when the residual dependence upon $E_h$ is linear
in $1/E_h$ \cite{Guagnelli:2002jd}. 
In the case of a flavored meson where $E_h$ is the heavy quark
mass, this amounts to state that a heavy quark effective theory expansion
is valid on finite volumes.
Hence, eq.~(\ref{eq:scaling}) has to be corrected as
\beq
\sigma_{\mathcal O}(E_\ell,E_h,L) = \sigma_{\mathcal O}(E_\ell,L) + \frac{\alpha^{(1)}(E_\ell,L)}{E_h}
+ \frac{\alpha^{(2)}(E_\ell,L)}{E^2_h} + \dots
\label{eq:quasiscaling}
\eeq
The number of terms that have to be taken into account
in the numerical calculations depend upon the particular observable
(see sec.~\ref{sec:firststep}).
The ansatz (\ref{eq:quasiscaling}), for the particular case of
the heavy--light meson masses it is supported
by HQET, it has been numerically
checked and allows to
extrapolate the knowledge of the step scaling function 
to regions of phenomenological interest where direct simulations are
too expensive. 
The computation of the observable $\mathcal O$ proceeds according to the following lines. 

First,  $\mathcal O$ is computed on a small finite volume $L_0$, where the high
energy scale $E_h$ can match its phenomenological value
with the lattice cutoff much larger than $E_h$. 
This computation is clearly unphysical, because the finiteness of the volume produces a distortion
of the result, that cannot be compared with the experimental value. 

Second, the step scaling function is used 
in order to evolve this finite size measurement to a larger volume, according to 
\beq 
{\mathcal O}(E_\ell,E_h,L_\infty) = {\mathcal O}(E_\ell,E_h,L_0)\ \sigma_{\mathcal O}(E_\ell,E_h,L_0)\ \sigma_{\mathcal O}(E_\ell,E_h,2L_0)\dots
\label{eq:starting}
\eeq
In the large volumes the step scaling functions are evaluated at the high energy scale
$E_h$ by extrapolation, relying on the parametrization of eq.~(\ref{eq:quasiscaling}).

Each step of the calculation can be extrapolated to the continuum
limit. 
In order to match subsequent steps, a non--perturbative knowledge of the
lattice spacing $a(g_0)$ as function of the bare coupling is
required \cite{Guagnelli:1998ud,Necco:2001xg}; 
the range of validity of the non--perturbative parametrization of $a(g_0)$
has been extended to very small couplings by a renormalization group analysis
in \cite{Guagnelli:2002ia}. Throughout the paper we use $r_0=0.5$ fm.

The above strategy has been straightforwardly applied in this work to the case of the heavy--light meson masses, where the 
observable $\mathcal O$ is the meson mass $M$, extracted from non--perturbative simulations of two point 
correlators, and the energy scales $E_h$ and $E_\ell$ are the
heavy and the light quark masses 
($m_{h}$,$m_\ell$). 
The light quark mass is extrapolated to values corresponding to the 
physical $u$ and $s$ masses, while the heavy quark mass is determined by comparing the meson mass to the experimental
spectrum. 
The heavy--heavy mesons case should not present major finite
size effects, being characterized by a single heavy scale. However,
as we will see, the application of the SSM method allows to solve the
problem of the fast decay of the meson correlation functions.

\section{Renormalization}
\label{sec:framework}

The step scaling function is calculated within the SF
\cite{Luscher:1992an,Sint:1994un}, which has already been applied 
to a number of different 
finite size problems \cite{Luscher:1994gh,Capitani:1998mw,Bode:2001jv,Guagnelli:2003hw,Heitger:2003xg}. In particular, we use a topology $T\times L^3$ with 
periodic boundary conditions on the space directions, Dirichlet 
boundary conditions along time and the following set of parameters
\beq
T=2L, \qquad C = C' = 0, \qquad \theta = 0
\eeq
Here $C$ and $C'$ represent the boundary gauge fields and $\theta$ is a topological phase which affects the periodicity 
of the fermion boundary conditions. Lattice discretization is performed using non--perturbative $O(a)$ improved clover 
action  \cite{Luscher:1997ug} and operators. In order to set the notation, let 
\beqa
A_\mu(x) &=& \overline{\psi}_i(x) \gamma_\mu \gamma_5\psi_j(x) 
\nonumber \\ \nonumber \\
P(x) &=& \overline{\psi}_i(x) \gamma_5 \psi_j(x) 
\nonumber \\ \nonumber \\
V_\mu(x) &=& \overline{\psi}_i(x) \gamma_\mu \psi_j(x) 
\nonumber \\ \nonumber \\
T_{\mu\nu}(x) &=& \overline{\psi}_i(x) \gamma_\mu \gamma_\nu \psi_j(x) 
\eeqa
be the axial current, the axial density, the local vector current and the tensor bilinear operator 
respectively ($i$ and $j$ are flavor indices).
The improvement of the axial and vector currents is obtained through the relations
\beqa
A^I_\mu(x) &=& A_\mu(x) + a c_A \ \tilde{\partial}_\mu P(x)
\nonumber \\ \nonumber \\ 
V^I_\mu(x) &=& V_\mu(x) + a c_V \ \tilde{\partial}_\nu T_{\mu\nu}(x) 
\eeqa
where $\tilde{\partial}_\mu = (\partial_\mu + \partial_\mu^*)/2$ and $\partial_\mu$, $\partial^*_\mu$ are the usual 
forward and backward lattice derivatives respectively.
For what concerns the improvement coefficients, in the case of $c_A$ we use the non--perturbative results of \cite{Luscher:1997ug} while,
in the case of $c_V$, we use the non--perturbative data only at the values of the bare couplings where they exist \cite{Guagnelli:1998db}
and perturbative results otherwise \cite{Sint:1997jx}. 
The correlation functions used
to compute pseudoscalar and vector meson masses are defined by probing the previous operators with appropriate boundary quark sources  
\beqa
f^I_A(x_0) &=& -\frac{a^6}{2} \sum_{\bf y,z}\langle \overline{\zeta}_j({\bf y}) \gamma_5 \zeta_i({\bf z}) A^I_0(x) \rangle
\nonumber \\ \nonumber \\
f_P(x_0) &=& -\frac{a^6}{2} \sum_{\bf y,z}\langle \overline{\zeta}_j({\bf y}) \gamma_5 \zeta_i({\bf z}) P(x) \rangle \label{eq:correlations}
\\ \nonumber \\
f^I_V(x_0) &=& -\frac{a^6}{6} \sum_{\bf y,z}\langle \overline{\zeta}_j({\bf y}) \gamma_k \zeta_i({\bf z}) V_k(x) \rangle \nonumber
\eeqa
where $\zeta_i({\bf y})$ and $\overline{\zeta}_i({\bf y})$ can be considered as quark and anti--quark boundary 
states.

The so--called bare current quark masses are defined through the lattice version of the PCAC relation
\beq
m^{WI}_{ij} = \frac{ \tilde{\partial_0}f_A(x_0) + a c_A \partial_0^* \partial_0 f_P(x_0)  }{2 f_P(x_0)} 
\label{eq:nondiagonal}
\eeq
These masses are connected to the renormalization group invariant (RGI) quark masses, according to the definitions
given in \cite{Gasser:1985gg}, through a renormalization factor
which has been computed non--perturbatively in \cite{Capitani:1998mw}:
\beq
m_{ij}^{RGI} = Z_M(g_0) \ \left[ 1 + (b_A-b_P)\ \frac{am_i+am_j}{2} \right] \ m^{WI}_{ij}(g_0)
\label{eq:rgimassij}
\eeq
where $am_i$ is defined as
\beq
am_i = \frac{1}{2} \left[\frac{1}{k_i} - \frac{1}{k_c} \right] 
\label{eq:barequarkmass}
\eeq
The combination $b_A-b_P$ of the improvement coefficients of the axial current and pseudoscalar density has been
non--perturbatively computed in \cite{deDivitiis:1998ka,Guagnelli:2000jw}. The factor $Z_M(g_0)$ is known with very high precision in a 
range of inverse bare couplings that does not cover all the values of $\beta$ used in our simulations.
We have used the results reported in table~(6) of ref.~\cite{Capitani:1998mw} 
to parametrize $Z_M(g_0)$  in the enlarged range of $\beta$ values $\left(5.9,7.6\right)$.

The RGI mass of a given quark is obtained from eq.~(\ref{eq:rgimassij}) using the diagonal
correlations
\beq
m^{RGI}_i = m_{ii}^{RGI} 
\label{eq:rgimass1}
\eeq
From non--diagonal correlations in eq.~(\ref{eq:rgimassij}) 
one obtains different $O(a)$ improved definitions of the RGI $i$--quark mass 
for different choices of the $j$--flavor:
\beq
m^{RGI}_{i_{\{j\}}} = 2 m^{RGI}_{ij} - m^{RGI}_{jj} 
\label{eq:rgimassi}
\eeq
All these definitions must have the same continuum limit because the dependence upon the $j$--flavor 
is only a lattice artifact.
Further, for each definition we use in eq.~(\ref{eq:nondiagonal}) either standard lattice
time derivatives as well as improved ones \cite{deDivitiis:1998ka,Guagnelli:2000jw}.

Another non--perturbative $O(a)$ improved definition of the RGI quark masses can be obtained starting from the
bare quark mass 
\beq
\hat{m}_i^{RGI} = Z_M(g_0) \ Z(g_0) \ \left[ 1 + b_m\ am_i \right] \ m_i 
\label{eq:rgimass2}
\eeq
where the improvement coefficient $b_m$  
and the renormalization constant 
\beq
Z(g_0) = \frac{Z_m Z_P}{Z_A} 
\eeq 
have been non-perturbatively computed in ref.~\cite{deDivitiis:1998ka,Guagnelli:2000jw}.

Equations (\ref{eq:rgimass1}), (\ref{eq:rgimassi}) and (\ref{eq:rgimass2}) give us different possibilities 
to identify the valence quarks inside a given meson (fixed by the values of the bare quark masses).
The procedure is well defined on small volumes because the
RGI quark mass is a physical quantity that does not depend upon the scale,
given in the SF scheme by the volume, and is defined in terms of local correlations
that do not suffer finite volume effects.
Each pair $\left(m^{RGI}_i,m^{RGI}_j\right)$ fixed a priori is matched, changing
the values of the hopping parameters, 
by the different definitions of equations (\ref{eq:rgimass1}), (\ref{eq:rgimassi}) and (\ref{eq:rgimass2}),
and leads to values of the corresponding meson mass differing by $O(a^2)$ lattice artifacts.
We take advantage of this plethora of definitions by constraining  
in a single fit the continuum extrapolations
(see {\footnotesize {\bf Fig.} [{\bf \ref{fig:A-ContSV},\ref{fig:A-S1-Cont},\ref{fig:A-S2-Cont}}]}).

The meson masses are extracted from the so--called \emph{effective mass}
\beq
a M_X(x_0) = \frac{1}{2} \ln \left[ f_X(x_0 - a) / f_X(x_0 + a)\right]   
\label{eq:effmass}
\eeq
where $f_X$ is one of the correlations defined in (\ref{eq:correlations}).
On a physical volume, this quantity exhibits a plateau in the time region where
the ground state dominates the correlation and no boundary effects are present.
On a small volume the effective mass is affected by two different finite volume effects: the first one
is due to the compression of the low energy wavelength (if present) and the second one comes from
the presence of the excited states contribution to the correlation (no plateau).
A good definition of the meson mass that suits the step scaling method is obtained by choosing the value of the
effective mass at $x_0 = T/2$. 
In this case $T = 2L$ and 
the step scaling technique (see eq.(\ref{eq:starting})) 
connects $x_0 = L_{min}$, where the meson mass has been defined on the smallest volume,
with $x_0 = L_{max}$, where one expects to be free from both sources of finite volume effects.

\begin{figure}[t]
\begin{center}
\epsfig{file=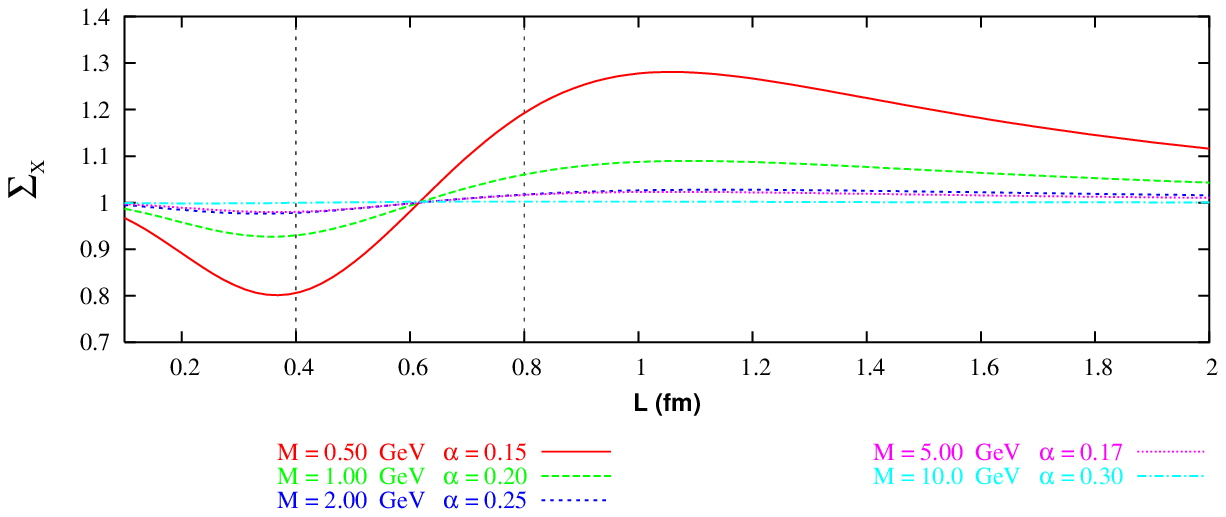,width=12cm}
\mycaption{The figure shows the excited states finite size effect as predicted by a simplified qualitative model discussed in the text.
The effective mass step scaling functions are plotted, at different values of the meson masses, as functions of the volume.}
\label{fig:model}
\end{center}
\end{figure}

The finite volume effects due to the excited states can be easily understood by means of a
\emph{two mass} model for the correlation
\beq
f_X(x_0,L) = A(L) \ e^{-M(L) x_0} + B(L)  \ e^{-\left[M(L)+\delta M(L)\right] x_0} 
\eeq
Here $A(L)$ and $B(L)$ are coefficients, $M(L)$ is the meson mass and $\delta M(L)$ is
the mass shift. All these quantities depend upon the physical extension of the volume $L$
but, in order to isolate the effects due to the excited states, we consider $M$ and $\delta M$ constants.
In this simplified model the effective mass of eq.~(\ref{eq:effmass})
takes the form
\beq
a M_X(x_0) = a M + \frac{1}{2} \ln \left\{ \frac{1 + \frac{B(L)}{A(L)} \ e^{-\delta M (x_0-a)} \ }{1 + \frac{B(L)}{A(L)} \ e^{-\delta M (x_0+a)}} \right\}
\eeq
As in numerical simulations, we take $x_0 = L$.
The functional dependence upon the volume of the ratio $B(L)/A(L)$ can
be inferred from the data
on the meson decay constants given in \cite{Guagnelli:2002jd,Guagnelli:2002adven}. 
A parametrization that fits well the data is given by
\beq
A(L) = \alpha \left( 1 + \frac{0.6}{L^2} \right) \qquad B(L) = -0.5 
\eeq   
In {\footnotesize {\bf Fig.} [{\bf \ref{fig:model}}]} are shown the plots of the
effective mass step scaling function, as derived from this model, defined by
\beq
\Sigma_X(L) = \frac{M_X(2L)}{M_X(L)} 
\label{eq:model}
\eeq
as functions of the volume and at different values of the meson masses. At the volume $L=0.4$ fm the step scaling
functions are smaller than one while,
the pattern is reversed at $L=0.8$ fm.
Independently from the volume, the heavier mesons are closer to unity than
the lighter ones. \\
The qualitative predictions of this simple model well reproduces the behavior of the numerically measured
step scaling functions, shown in {\footnotesize {\bf Fig.}~[{\bf \ref{fig:A-S1-sigmas}}]}
for $L=0.4$ fm and in {\footnotesize {\bf Fig.}~[{\bf \ref{fig:A-S2-sigmas}}]}
for $L=0.8$ fm.

\section{Step scaling functions and HQET}
\label{sec:hqet}

In this section we enter into the details
of the application of the SSM, illustrated in sec.~\ref{sec:stepscalingmethod},
to the particular case of the heavy--light meson masses computation.
We introduce the step scaling functions
\beqa
\sigma_P \left(L, m^{RGI}_1, m^{RGI}_2 \right) = \frac{\left.M_P\left(m^{RGI}_1, m^{RGI}_2 \right)\right|_{2L}}
{\left.M_P\left(m^{RGI}_1, m^{RGI}_2 \right)\right|_{L}}
\nonumber \\ \nonumber \\
\sigma_V \left(L, m^{RGI}_1, m^{RGI}_2 \right) = \frac{\left.M_V\left(m^{RGI}_1, m^{RGI}_2 \right)\right|_{2L}}
{\left.M_V\left(m^{RGI}_1, m^{RGI}_2 \right)\right|_{L}}
\label{eq:masssigmas}
\eeqa
for the pseudoscalar and vector meson masses respectively. \\
To validate the
expansion of eq.~(\ref{eq:quasiscaling}), we can make use
of the HQET predictions on the heavy--light meson masses.
In the infinite volume the pseudoscalar and vector
meson masses have the following expansion in terms of the heavy--quark
mass:
\beq
M_X(m_h, m_l) = m_h + \bar{\Lambda}(m_l) + \frac{\alpha_X(m_l)}{m_h} + \dots
\label{eq:hqetiv}
\eeq 
where $X \in \{P,V\}$.
Assuming the contribution of the $1/m^2_h$ corrections to be negligible,
at finite volume one has
\beq
M_X(m_h, m_l, L) = m_h + \bar{\Lambda}_X(m_l,L) + \frac{\alpha_X(m_l,L)}{m_h} + \dots
\label{eq:hqetsv}
\eeq 
where $\bar{\Lambda}_X(m_l,L)$ depends
upon the spin of the meson state because of the contamination
of the excited states to the finite volume correlations.
Using eqs.~(\ref{eq:masssigmas}) and (\ref{eq:hqetsv})
we obtain the HQET predictions for the step
scaling functions of the heavy-light
meson masses
\beq
\Sigma_X(L,m_h,m_l) = 1 + \frac{\Sigma^{(0)}_X(m_l,L)}{m_h} 
+ \frac{\Sigma^{(1)}_X(m_l,L)}{m_h^2} + \dots
\label{eq:sigmahqetsv}
\eeq
This result requires some considerations.
First we want to stress that in the infinite heavy--quark mass
limit the step scaling functions have to be exactly equal to one, 
$\Sigma_X(L, m_l, m_h \rightarrow \infty) = 1$\footnote{We thank A.~Kronfeld 
and R.~Sommer for having pointed out this property of $\Sigma_X$.}. 
This represents a strong constraint for the fits of the
heavy--quark mass dependence of the step scaling functions.

The second observation concerns the number of terms to be considered in eq.~(\ref{eq:sigmahqetsv}).
At order $O(1/m_h)$ one has
\beq
\Sigma^{(0)}_X(m_l,L) = \bar{\Lambda}_X(m_l,2L) - \bar{\Lambda}_X(m_l,L)
\eeq
corresponding to the static approximation in
eq.~(\ref{eq:hqetsv}). 
By increasing the physical volume $L$, the difference between
$\bar{\Lambda}_X(m_l,2L)$ and $\bar{\Lambda}_X(m_l,L)$
decreases because the two quantity have to be equal
in the infinite volume limit, making
the heavy--quark mass expansion of the finite volume effects
rapidly convergent. \\
The same arguments apply to the coefficient $\Sigma^{(1)}_X(m_l,L)$ that has to
be considered when in the expansion of the meson masses, eq.~(\ref{eq:hqetsv}),
the order $O(1/m_h)$ is taken into account.\\
In our calculation we will perform the fits of
the step scaling functions considering the $O(1/m_h^2)$
term, $\Sigma^{(1)}_X$, beyond the so called static evolution ({\it SE})
that retains $\Sigma^{(0)}_X$ only.
We will also report for comparison the fits for the {\it SE}
that are anyway compatible. 

Similar arguments could be repeated for
the heavy--heavy step scaling functions using
NRQCD. In this case one cannot predict the
value of the step scaling functions at $m_h \rightarrow \infty$.
With the three simulated values of the heavy quark masses
we extrapolate our numerical data
with a simplified
NLO formula, i.e. a linear dependence upon the heavy--quark mass, 
given the little sensitivity to
possible slowly varying logarithmic terms.

\section{Numerical results}
\label{sec:numresults}

This section contains the simulation parameters and the numerical results
for each step of the calculation.

\begin{table}[t]
\begin{center}
\scriptsize
\begin{tabular}{cccccc}
\midrule
$\beta$ & $L/a$         & $k_c$         & $k$      & $\qquad$ $m^{RGI}$ (GeV) $\qquad$ \\
\toprule
      &                 &               &  0.120081 &    7.14(8)    \\
      &                 &               &  0.120988 &    6.63(7)    \\
      &                 &               &  0.126050 &    4.024(44)  \\
6.963 & $16$            & 0.134827(6)   &  0.131082 &    1.696(19)  \\
      &                 &               &  0.131314 &    1.591(18)  \\
      &                 &               &  0.134526 &    0.1381(30) \\
      &                 &               &  0.134614 &    0.0978(28) \\
      &			&	        &  0.134702 &    0.0574(28) \\
\midrule			             
      &                 &               &  0.124176 &    7.11(8)    \\
      &                 &               &  0.124844 &    6.61(20)   \\
      &                 &               &  0.128440 &    4.018(44)  \\
7.300 & $24$            & 0.134235(3)   &  0.131800 &    1.695(19)  \\
      &                 &               &  0.131950 &    1.592(18)  \\
      &                 &               &  0.134041 &    0.1374(27) \\
      &                 &               &  0.134098 &    0.0971(24) \\
      &                 &               &  0.134155 &    0.0567(24) \\
\midrule							             
      &                 &               &  0.126352 &    7.10(8)    \\
      &                 &               &  0.126866 &    6.60(7)    \\
      &                 &               &  0.129585 &    4.016(44)  \\
7.548 & $32$            & 0.133838(2)   &  0.132053 &    1.698(19)  \\
      &                 &               &  0.132162 &    1.595(18)  \\
      &                 &               &  0.133690 &    0.1422(27) \\
      &                 &               &  0.133732 &    0.1021(25) \\
      &                 &               &  0.133773 &    0.0618(23) \\
\bottomrule
\end{tabular}
\mycaption{Simulation parameters  at $L_0 = 0.4$ fm. The RGI quark masses
are obtained using eq.~(\ref{eq:rgimass1}).}
\label{tab:SVSimPar}
\end{center}
\end{table}

\subsection{The small volume: $L_0 = 0.4$ fm}
\label{sec:smallvolume}

\begin{figure}[t]
\begin{center}
\epsfig{file=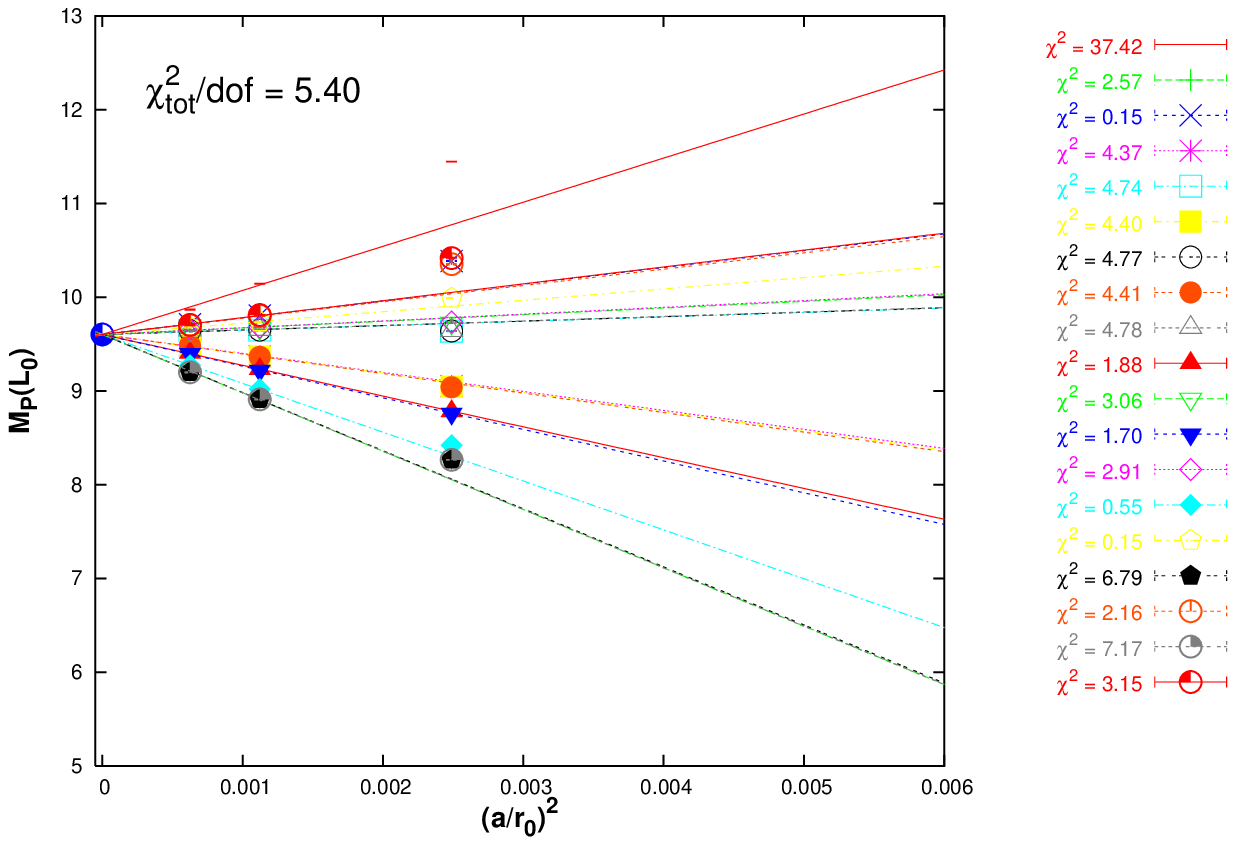,width=\textwidth}
\mycaption{Continuum extrapolation on the small volume, $L_0$, of the mass $M_P(L_0)$
of the pseudoscalar heavy--heavy meson
corresponding to the heavy quark of mass $m^{RGI} = 7.10$ GeV.
The different values of the meson mass correspond to different definitions
of the RGI quark masses given in equations
(\ref{eq:rgimass1}), (\ref{eq:rgimassi}) and (\ref{eq:rgimass2}). 
Units are in GeV.
Similar plots can be obtained, from the data reported in 
{\footnotesize {\bf Table} [{\bf\ref{tab:SVNumRes},\ref{tab:SVNumRes2},\ref{tab:SVNumRes3}}]}, 
for the other 
combinations of quark masses used in our simulations, also in the case of $M_V(L_0)$.\label{fig:A-ContSV}}
\end{center}
\end{figure}

The physical extension of the small volume has been chosen in order to properly account
the dynamics of quarks with masses in the region of the the physical $b$--quark;
we have fixed it to be $L_0 = 0.4$ fm. In order to have a continuum extrapolation of the
numerical results, this volume has been simulated
using three different discretization, $32\times 16^3$, $48\times 24^3$ and $64\times 32^3$.
Using the results reported in \cite{PDBook} we have fixed two heavy
quark masses with values interpolating the $b$--quark mass.
In the renormalization group invariant (RGI) scheme fixed by the conventions reported in
\cite{Gasser:1985gg,Capitani:1998mw}, these masses are $7.10$ GeV and $6.60$ GeV respectively.
Two other heavy quarks have been simulated in order to interpolate the mass region where, using the 
results reported in \cite{Rolf:2002gu,PDBook}, one expects to find the physical
$c$--quark, i.e. $1.70$ GeV and $1.60$ GeV respectively. The charm quark mass is affordable
also by a direct computation: here the calculation via our step scaling method represents a 
check of the procedure.
An additional heavy quark has been simulated with mass $4.00$ GeV.
Three light quark have been simulated with masses of $0.14$ GeV, $0.10$ GeV and
$0.06$ GeV. Using  the accurate determination of the RGI strange quark mass given in \cite{Garden:1999fg}
we have fixed one of the simulated light quarks to be the physical $s$. 
All the parameters of the three different simulations are summarized in \tab{tab:SVSimPar}. 
The numerical results of the pseudoscalar and vector meson masses, $M_P$ and $M_V$,  
are shown in 
{\footnotesize {\bf Table} [{\bf\ref{tab:SVNumRes},\ref{tab:SVNumRes2},\ref{tab:SVNumRes3}}]} both for the 
heavy--heavy and for the heavy--light quark anti--quark pairs.

\begin{table}[t]
\begin{center}
\scriptsize
\begin{tabular}{cccccc}
\midrule
$\beta$ & $L_0/a$         & $k_c$       & $k$      & $\qquad$ $m^{RGI}$ (GeV) $\qquad$\\
\toprule
      &                 &               & 0.120674 &  3.543(39)  \\
      &                 &               & 0.122220 &  3.114(34)  \\
      &                 &               & 0.126937 &  1.927(21)  \\
6.420 & $8$             & 0.135703(9)   & 0.134304 &  0.3007(36) \\
      &                 &               & 0.134770 &  0.2003(28) \\
      &                 &               & 0.135221 &  0.1028(21) \\
\midrule
      &                 &               & 0.1249   &  3.542(39)  \\
      &                 &               & 0.1260   &  3.136(34)  \\
      &                 &               & 0.1293   &  1.979(22)  \\
6.737 & $12$            & 0.135235(5)   & 0.1343   &  0.3127(38) \\
      &                 &               & 0.1346   &  0.2090(28) \\
      &                 &               & 0.1349   &  0.1080(21) \\
\midrule
      &                 &               & 0.127074 &  3.549(39)  \\
      &                 &               & 0.127913 &  3.153(35)  \\
      &                 &               & 0.130409 &  2.003(22)  \\
6.963 & $16$            &  0.134832(4)  & 0.134145 &  0.3134(38) \\
      &                 &               & 0.134369 &  0.2112(28) \\
      &                 &               & 0.134593 &  0.1086(20) \\
\bottomrule
\end{tabular}
\mycaption{Simulation parameters for the first evolution step $L_0 \to L_1 = 0.8$ fm.
The RGI quark masses are obtained using eq.~(\ref{eq:rgimass1}).}
\label{tab:S1simpar}
\end{center}
\end{table}

In order to obtain physical predictions from the simulated data on this finite volume 
we need to extrapolate our numerical results to the continuum.
As already mentioned, we have obtained different set of data
by using the different definitions of the RGI quark masses
given in the equations (\ref{eq:rgimass1}), (\ref{eq:rgimassi}) and (\ref{eq:rgimass2}). 
The continuum results are thus obtained trough a combined fit of all the set of data,  linear in $(a / r_0 )^2$, 
as shown in {\footnotesize {\bf Fig.} [{\bf \ref{fig:A-ContSV}}]} in the case of the mass
of the pseudoscalar heavy--heavy meson corresponding to the heavy quark of mass
$m^{RGI} = 7.10$ GeV.
We obtain a global $\chi^2/dof = 5.40$ to be compared with the $\chi^2$s of each individual definition
listed in the figure. We also show the points at the largest lattice spacing not included
in the fit. \\
The errors included in the evaluation of the $\chi^2$ are statistical only. These
are calculated by a jackknife procedure, also in the case of the step scaling functions.
The systematics due to the uncertainty on the lattice spacing
has been estimated by repeating the fit using the different values of the scale
allowed by the uncertainties quoted in \cite{Guagnelli:1998ud,Necco:2001xg,Guagnelli:2002ia}
and considering the spread of the results.
The same procedure has been used for the systematics due to the uncertainties on
the renormalization constants.
The resulting $2\%$ percent for the renormalization
constants and $1\%$ percent for the scale are
summed in quadrature and added to the statistical errors.

\subsection{The first evolution step: $L_0 \mapsto L_1 = 0.8$ fm}
\label{sec:firststep}

The finite volume effects on the quantities calculated at $L_0$,
are measured doubling the volume, $L_1 = 0.8$ fm, by using the 
step scaling functions of eqs.~(\ref{eq:masssigmas}).\\
In order to have results in the continuum, also in the case of the step scaling functions 
three different discretizations  of $L_0$
have been used, i.e $16\times 8^3$, $24\times 12^3$ and $32\times 16^3$. The volume $L_1$ has been 
simulated starting from the discretizations of $L_0$, fixing the value of the bare coupling
and doubling the number of lattice points in each direction.

\begin{figure}[htp]
\begin{center}
\epsfig{file=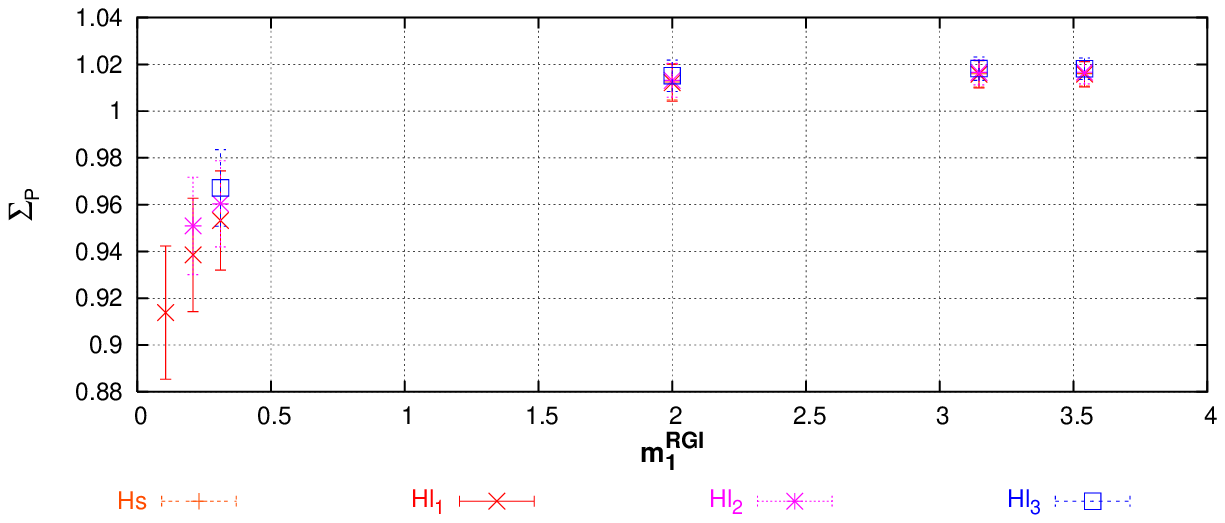,width=10cm}
\mycaption{
The figure shows the pseudoscalar step scaling functions $\Sigma_P$ as functions of $m_1^{RGI}$,
for the simulation of the first evolution step
corresponding to $\beta = 6.963$.
The different sets of data correspond to the values of $m_2^{RGI}$.
As can be seen
the step scaling functions approach a plateau for high values of $m_1^{RGI}$. 
Similar plots can be obtained using the data of 
{\footnotesize {\bf Table} [{\bf\ref{tab:S1NumRes1},\ref{tab:S1NumRes2},\ref{tab:S1NumRes3}}]}
for the other values of the bare coupling and for the vector mesons step scaling
functions $\Sigma_V$.}
\label{fig:A-S1-sigmas}
\end{center}
\end{figure}

\begin{figure}[htp]
\begin{center}
\epsfig{file=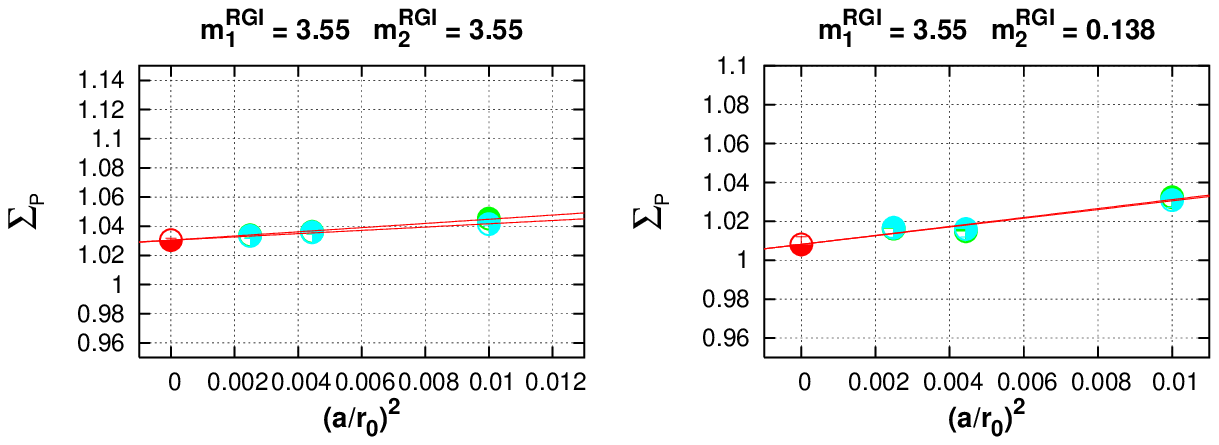,width=\textwidth}
\mycaption{Continuum extrapolation on the first evolution step, $L_0\mapsto L_1$, of the 
step scaling function, $\Sigma_P(L_0)$,
of the pseudoscalar meson
corresponding to the heavy quark of mass $m_1^{RGI} = 3.55$ GeV.
The two sets of data are obtained using the two definitions of RGI quark masses of equations
(\ref{eq:rgimass1}) and (\ref{eq:rgimass2}). 
Units are in GeV.
Similar plots can be obtained, from the data reported in 
{\footnotesize {\bf Table} [{\bf\ref{tab:S1NumRes1},\ref{tab:S1NumRes2},\ref{tab:S1NumRes3}}]}, for the other 
combinations of quark masses used in our simulations, also in the case of $\Sigma_V(L_0)$.\label{fig:A-S1-Cont}}
\end{center}
\end{figure}

\begin{figure}[htp]
\begin{center}
\epsfig{file=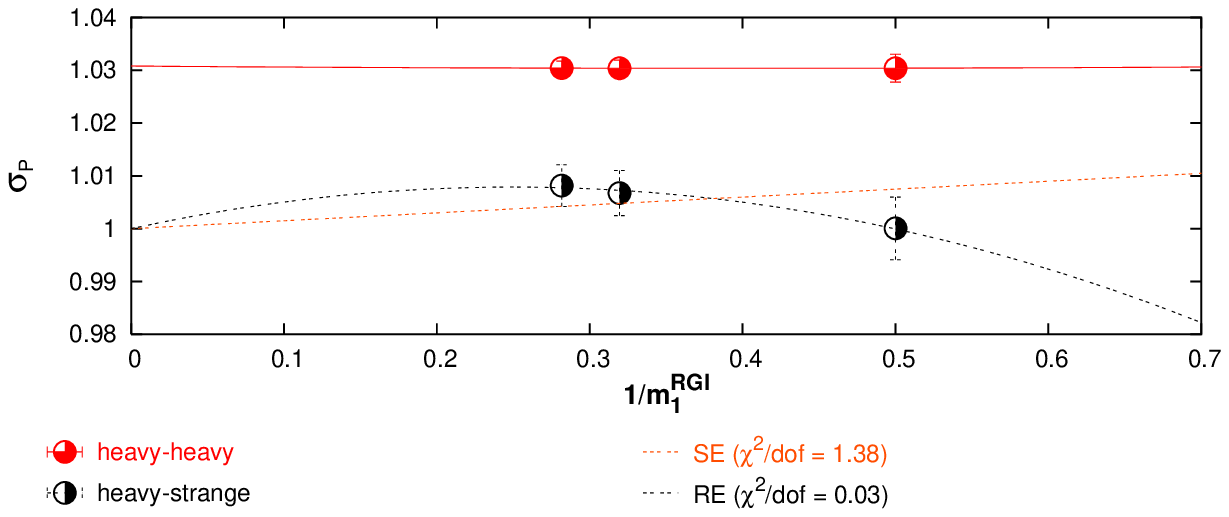,width=10cm}
\mycaption{
The figure shows the continuum extrapolated step scaling functions $\sigma_P(L_0)$
as functions of $1/m_1^{RGI}$.
The heavy extrapolations are shown only for the heavy--strange (Hs) set of data.
Similar plots can be obtained using the data of 
{\footnotesize {\bf Table} [{\bf\ref{tab:S1NumRes1},\ref{tab:S1NumRes2},\ref{tab:S1NumRes3}}]}
for the vector mesons step scaling
functions $\sigma_V(L_0)$.}
\label{fig:A-S1-sigmas2}
\end{center}
\end{figure}

The simulated quark masses have been halved with respect to the masses simulated on the small volume
in order to have the same order of discretization effects proportional to $am$.
The set of parameters for the simulations of this evolution step is reported in \tab{tab:S1simpar} and
the numerical results are given in {\footnotesize {\bf Table} [{\bf\ref{tab:S1NumRes1},\ref{tab:S1NumRes2},\ref{tab:S1NumRes3}}]}. 

The step scaling functions of the pseudoscalar mesons at $\beta=6.963$ are plotted, at fixed $m^{RGI}_2$, as 
functions of $m^{RGI}_1$ in {\footnotesize {\bf Fig.} [{\bf \ref{fig:A-S1-sigmas}}]}.
The value of the step scaling functions for the $s$ quark are obtained trough linear interpolation.

Both $\Sigma_P$ and $\Sigma_V$ are almost flat in a region of
heavy quark masses starting around the charm mass.
The hypothesis of low sensitivity upon the high--energy scale is thus verified. \\
In {\footnotesize {\bf Fig.} [{\bf \ref{fig:A-S1-Cont}}]} are reported 
the results of the continuum extrapolation of the 
step scaling function, $\Sigma_P(L_0)$,
of the pseudoscalar meson
corresponding the heavy quark of mass $m_1^{RGI} = 3.55$ GeV. \\
The residual heavy mass dependence of the continuum extrapolated step
scaling functions 
is very mild, as shown in {\footnotesize {\bf Fig.} [{\bf \ref{fig:A-S1-sigmas2}}]}
in the plot of $\sigma_P$ as a function of the inverse quark mass. \\
In the heavy--light case, as already discussed in sec.~\ref{sec:hqet}, we extract the values of
the step scaling functions at the values of the heavy--quark masses
simulated on the small volume by interpolation between the numerical
results and the theoretical point at $m_h^{RGI} \rightarrow \infty$
using both a linear fit ({\it SE}) and a quadratic fit ({\it RE}).
In the heavy--heavy case the results are linearly extrapolated.

\begin{table}[t]
\begin{center}
\scriptsize
\begin{tabular}{cccccc}
\midrule
$\beta$ & $L_1/a$       & $k_c$         & $k$      & $\qquad$ $m^{RGI}$ (GeV) $\qquad$ \\
\toprule
      &                 &               & 0.118128 &  2.012(22)  \\
      &                 &               & 0.121012 &  1.551(17)  \\
      &                 &               & 0.122513 &  1.337(15)  \\
5.960 & $8$             & 0.13490(4)    & 0.131457 &  0.3154(36) \\
      &                 &               & 0.132335 &  0.2322(28) \\
      &                 &               & 0.133226 &  0.1466(44) \\
\midrule
      &                 &               & 0.124090 &  1.984(22)  \\
      &                 &               & 0.126198 &  1.584(17)  \\
      &                 &               & 0.127280 &  1.389(15)  \\
6.211 & $12$            & 0.135831(8)   & 0.133574 &  0.3493(39)  \\
      &                 &               & 0.134177 &  0.2550(29)  \\
      &                 &               & 0.134786 &  0.1510(19) \\
\midrule
      &                 &               & 0.126996 &  1.933(21)  \\
      &                 &               & 0.128646 &  1.547(17) \\
      &                 &               & 0.129487 &  1.355(14) \\
6.420 & $16$            &  0.135734(5)  & 0.134318 &  0.3016(34) \\
      &                 &               & 0.134775 &  0.2038(24) \\
      &                 &               & 0.135235 &  0.1055(15) \\
\bottomrule
\end{tabular}
\mycaption{Simulation parameters for the first evolution step $L_1 \to L_2 = 1.6$ fm.
The RGI quark masses are obtained using eq.~(\ref{eq:rgimass1}).}
\label{tab:S2simpar}
\end{center}
\end{table}

\subsection{The second evolution step: $L_1 \mapsto L_2 = 1.6$ fm}
\label{sec:secondstep}

\begin{figure}[htp]
\begin{center}
\epsfig{file=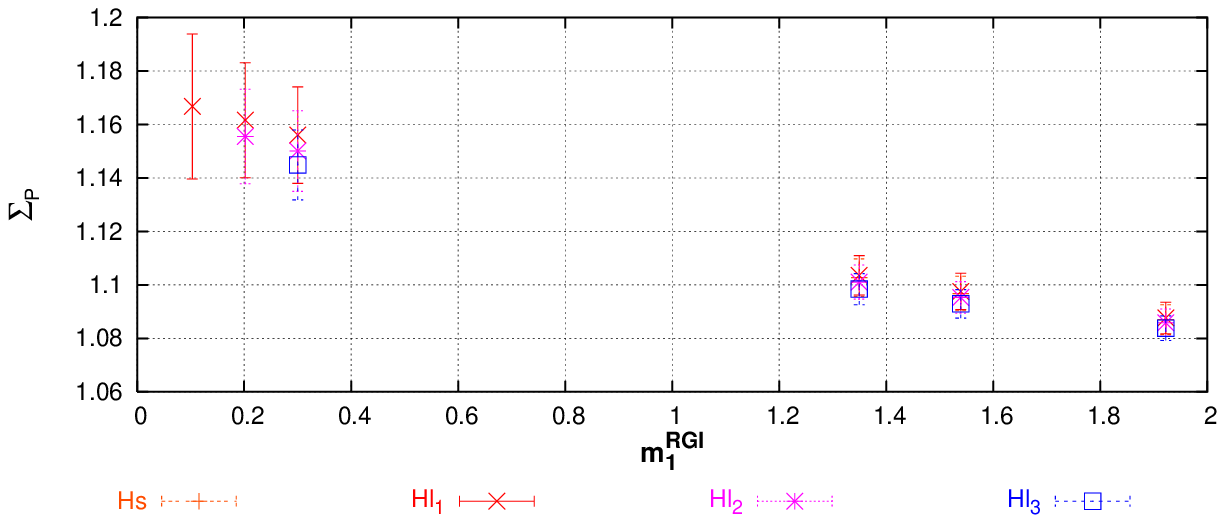,width=10cm}
\mycaption{
The figure shows the pseudoscalar step scaling functions $\Sigma_P$ as functions of $m_1^{RGI}$,
for the simulation of the second evolution step
corresponding to $\beta = 6.420$.
The different sets of data correspond to the values of $m_2^{RGI}$.
Similar graphs can be obtained using the data of 
{\footnotesize {\bf Table} [{\bf\ref{tab:S2NumRes1},\ref{tab:S2NumRes2},\ref{tab:S2NumRes3}}]}
for the other values of the bare coupling and for the vector mesons step scaling
functions $\Sigma_V$.}
\label{fig:A-S2-sigmas}
\end{center}
\end{figure}

\begin{figure}[htp]
\begin{center}
\epsfig{file=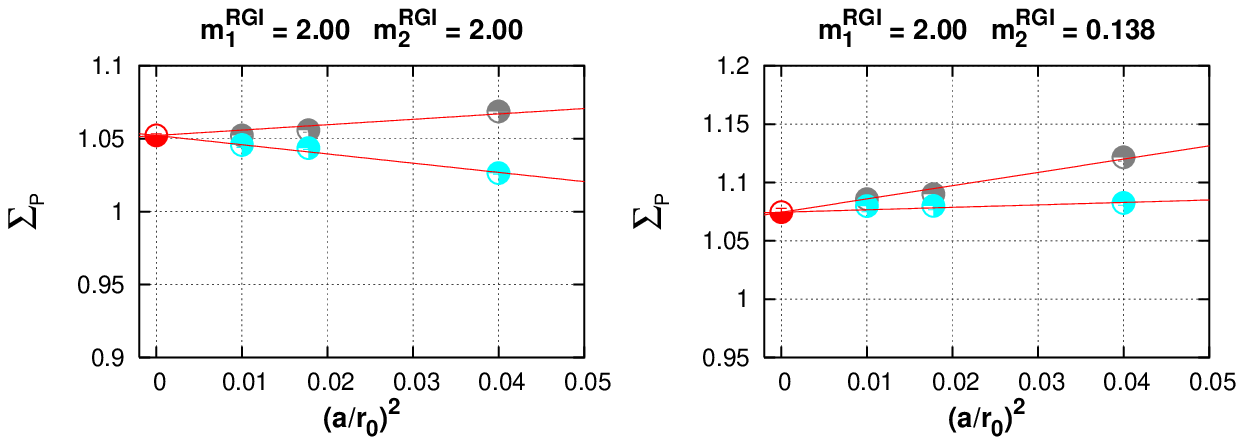,width=\textwidth}
\mycaption{Continuum extrapolation on the second evolution step, $L_1\mapsto L_2$, of the 
step scaling function, $\Sigma_P(L_1)$,
of the pseudoscalar meson
corresponding to the heavy quark of mass $m_1^{RGI} = 2.00$ GeV.
The two sets of data are obtained using the two definitions of RGI quark masses of equations
(\ref{eq:rgimass1}) and (\ref{eq:rgimass2}). 
Units are in GeV.
Similar plots can be obtained, from the data reported in 
{\footnotesize {\bf Table} [{\bf\ref{tab:S2NumRes1},\ref{tab:S2NumRes2},\ref{tab:S2NumRes3}}]}, for the other 
combinations of quark masses used in our simulations, also in the case of $\Sigma_V(L_1)$.\label{fig:A-S2-Cont}}
\end{center}
\end{figure}

\begin{figure}[htp]
\begin{center}
\epsfig{file=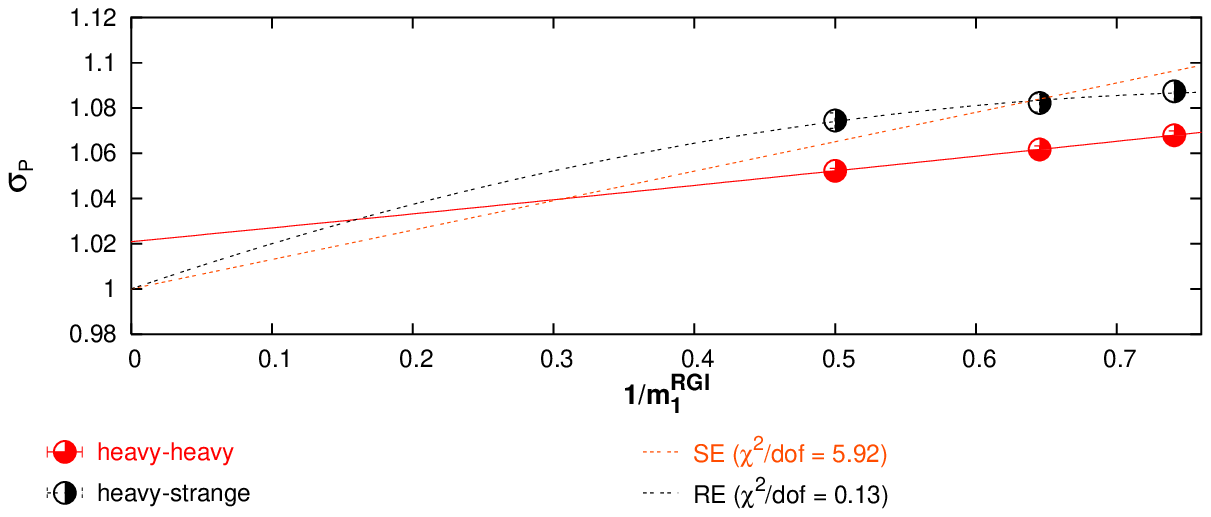,width=10cm}
\mycaption{
The figure shows the continuum extrapolated step scaling functions $\sigma_P(L_1)$
as functions of $1/m_1^{RGI}$.
The heavy extrapolations are shown only for the heavy--strange (Hs) set of data.
Similar plots can be obtained using the data of 
{\footnotesize {\bf Table} [{\bf\ref{tab:S2NumRes1},\ref{tab:S2NumRes2},\ref{tab:S2NumRes3}}]}
for the vector mesons step scaling
functions $\sigma_V(L_1)$.}
\label{fig:A-S2-sigmas2}
\end{center}
\end{figure}

In order to have the results on a physical volume, $L_2 = 1.6$ fm, a second evolution step
it is required. This is done computing the meson mass step scaling functions
of eq.~(\ref{eq:masssigmas}) at $L=L_1$, by the procedure outlined in the
previous section. The parameters of the simulations are given in \tab{tab:S2simpar}
and the results are in {\footnotesize {\bf Table} [{\bf\ref{tab:S2NumRes1},\ref{tab:S2NumRes2},\ref{tab:S2NumRes3}}]}.

Also in this case, the values of the simulated quark masses have been halved with respect to
the previous step, owing to the lower values of the simulation cutoffs.
In {\footnotesize {\bf Fig.}~[{\bf \ref{fig:A-S2-sigmas}]} we
show the pseudoscalar step scaling functions at $\beta = 6.420$
and in {\footnotesize {\bf Fig.}~[{\bf \ref{fig:A-S2-sigmas2}]}
the residual heavy--quark mass dependence with the {\it SE} and {\it RE}
fits.

{\footnotesize {\bf Fig.} [{\bf \ref{fig:A-S2-Cont}}]} shows
the continuum extrapolation  of the 
step scaling function, $\Sigma_P(L_1)$,
of the pseudoscalar meson
corresponding to the heavy quark of mass $m_1^{RGI} = 2.00$ GeV.

The contributions of the excited states, predicted from the simple model of eq.~(\ref{eq:model}) and
present in our numerical data, should 
disappear in a third evolution step.
A check supporting this hypothesis is that on the larger volumes used in the simulation of this
evolution step, i.e. $L_2$ at $\beta=6.420$ and $\beta=6.211$, the values
of the meson masses defined at $x_0=T/2$ coincide, within the errors, with the
values coming from a \emph{single mass} fit to the correlations.

\section{Physical results}
\label{sec:physresults}

\begin{table}[t]
\begin{center}
\scriptsize
\begin{tabular}{cccc}
\midrule
$\qquad m_h^{RGI} \qquad$ & $\qquad$ State $\qquad$      & $\qquad M_P \qquad$         & $\qquad M_V \qquad$     \\
\toprule
      &    $\bar{h}h$     &   10.11(22)   &  10.12(22) \\
7.10  &    $\bar{h}s$     &   5.48(13)    &  5.52(13)\\
      &    $\bar{h}u$     &   5.40(16)    &  5.44(16)\\
\midrule
      &    $\bar{h}h$     &   9.49(21)    &  9.50(21) \\
6.60  &    $\bar{h}s$     &   5.18(12)    &  5.22(12)\\
      &    $\bar{h}u$     &   5.10(15)    &  5.14(15)\\
\midrule
      &    $\bar{h}h$     &   6.18(14)    &  6.20(14) \\
4.00  &    $\bar{h}s$     &   3.55(8)     &  3.62(8) \\
      &    $\bar{h}u$     &   3.46(10)    &  3.53(11)\\
\midrule
      &    $\bar{h}h$     &   3.15(7)     &  3.21(7)\\
1.70  &    $\bar{h}s$     &   1.97(5)     &  2.09(5)\\
      &    $\bar{h}u$     &   1.88(6)     &  2.00(6)\\
\midrule
      &    $\bar{h}h$     &   3.01(7)     &  3.07(7)\\
1.60  &    $\bar{h}s$     &   1.90(5)     &  2.01(5)\\
      &    $\bar{h}u$     &   1.80(5)     &  1.93(6)\\
\bottomrule
\end{tabular}
\mycaption{Meson masses in the infinite volume limit. Units are in GeV.}
\label{tab:physnum}
\end{center}
\end{table}

\begin{figure}
\begin{center}
\epsfig{file=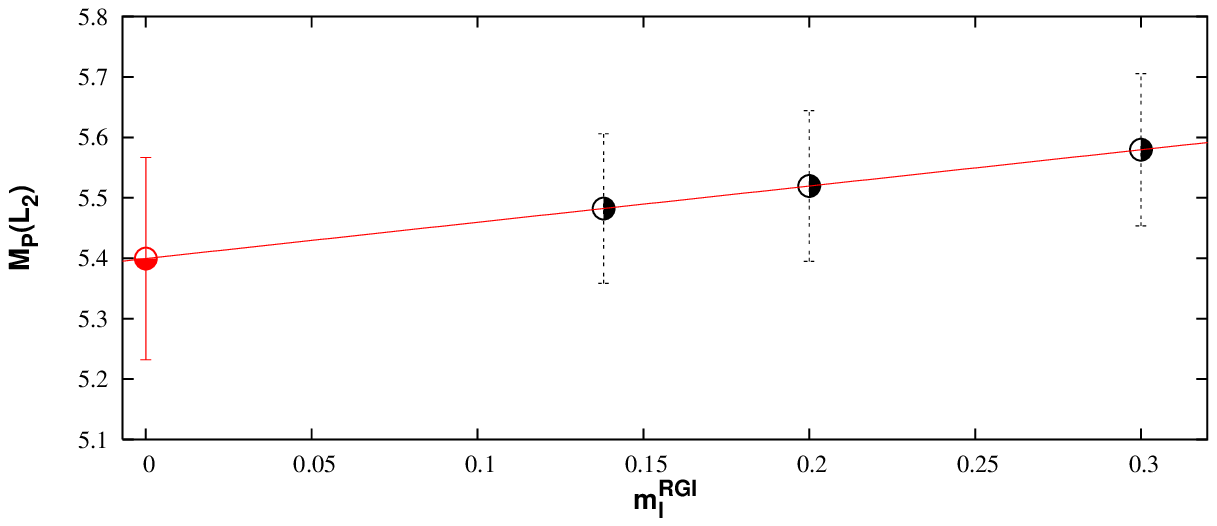,width=10cm}
\mycaption{
Chiral extrapolation of the continuum results for the
pseudoscalar meson corresponding to $m_h^{RGI} = 7.1$.
Units are in GeV.
}
\label{fig:chirallarge}
\end{center}
\end{figure}

In this section we combine the results of the small
volume with the results of the step scaling functions
to obtain, according to eq.~(\ref{eq:starting}), the physical numbers.
In {\footnotesize {\bf Table} [{\bf\ref{tab:physnum}}]} we give
the pseudoscalar and vector meson masses corresponding to
the heavy quarks simulated on the small volume and, in the
heavy--light case, to strange and up light quarks.
The results corresponding to the up quark have been obtained
by extrapolating the continuum data in the
large volume from masses around the strange region.
In {\footnotesize {\bf Figure} [{\bf\ref{fig:chirallarge}}]}
we show the extrapolation for the pseudoscalar meson
corresponding to $m_h^{RGI} = 7.1$ GeV. Having
simulated three light quark on the largest volume we have
fitted the results linearly without trying 
complicated functional forms requiring more than two
parameters.
Comparing these results with the experimental determinations
of the same quantities \cite{PDBook} we obtain
different determinations of the $b$--quark mass, depending upon
the physical state used as experimental input.
The results are summarized
in {\footnotesize {\bf Table} [{\bf\ref{tab:physnumQM}}]}.

\begin{table}[h]
\begin{center}
\scriptsize
\begin{tabular}{ccc}
\midrule
$\qquad$ State $\qquad$  & \multicolumn{2}{c}{$m^{RGI}_h$}  \\
\midrule   
             & $\qquad\qquad$ from P  $\qquad\qquad$ &  $\qquad\qquad$ from V  $\qquad\qquad$ \\
\toprule
$\bar{b}b$    & 6.44(14)              &  6.57(14)              \\
$\bar{b}s$    & 6.91(16)              &  6.93(16)              \\
$\bar{b}u$    & 6.90(20)              &  6.91(20)              \\
\midrule
$\bar{c}c$    & 1.603(35)             &  1.642(36)             \\
$\bar{c}s$    & 1.692(38)             &  1.741(39)             \\
$\bar{c}u$    & 1.690(50)             &  1.712(51)             \\
\bottomrule
\end{tabular}
\mycaption{Determinations of the heavy quark masses from the heavy--heavy and from
the heavy--light states. The errors include our estimate of the systematics. Units are in GeV.}
\label{tab:physnumQM}
\end{center}
\end{table}

Within the quenched approximation, the determinations of the
quark masses coming from the heavy--heavy or from the heavy--light spectrum
in principle differ because the theory does not account for the fermion loops. 
We obtain two determinations that are marginally compatible within the errors and that
might suggest the need for a tiny unquenching effect.
The good agreement between the determinations of the quark masses coming from
the heavy--up and heavy--strange sets of data make us confident on our
chiral extrapolations.
Of course the heavy--strange and the heavy--heavy case are not extrapolated at all
and we therefore use these results only to determine the heavy--quark masses.
The numbers we quote as final results are obtained by averaging the four results 
in the first two rows of the two sets of 
{\footnotesize {\bf Table} [{\bf\ref{tab:physnumQM}}]} for the heavy-heavy and the heavy-strange cases 
and by keeping the typical error of a single case:
\beq
m_b^{RGI} = 6.73(16) \mbox{ GeV} \qquad \qquad m_b^{\overline{MS}}(m_b^{\overline{MS}}) = 4.33(10) \mbox{ GeV}
\label{eq:mbfinal}
\eeq
for the $b$--quark and
\beq
m_c^{RGI} = 1.681(36) \mbox{ GeV} \qquad \qquad m_c^{\overline{MS}}(m_c^{\overline{MS}}) = 1.319(28) \mbox{ GeV}
\label{eq:mcfinal}
\eeq
for the charm. 
The latter results compare favorably with the results of the direct computations \cite{Rolf:2002gu,PDBook}.

As already explained in section~\ref{sec:smallvolume}, our error estimate includes both the \emph{statistical} error from the Monte Carlo
simulation as well as the \emph{systematic} error coming from the uncertainty on the lattice spacing corresponding at a given $\beta$ value
and to the uncertainty on the renormalization constants in eqs~(\ref{eq:rgimassij}) and~(\ref{eq:rgimass2}).
The final errors on the continuum quantities, of the order of $2\%$ percent for the renormalization
constants and of about $1\%$ percent for the scale, are
summed in quadrature and added to the statistical errors.
The evolution to the $\overline{MS}$ scheme has been done
using four--loop renormalization group equations \cite{vanRitbergen:1997va,Vermaseren:1997fq,Chetyrkin:1997dh}.

Using these determinations of the quark masses we can calculate
the mass of the $B_c$ meson from our set of correlations suitably interpolated.
Following this procedure we obtain a determination
that is in very good agreement with the experimental measurement of the meson mass \cite{PDBook}:
\beq
M_{B_c} = 6.46(15) \mbox{ GeV} \qquad  M_{B_c}^{exp} = 6.40(39)(13) \mbox{ GeV}
\label{eq:mbcfinal}
\eeq
The same analysis has been done also in the case of the decay constants and in 
\cite{Guagnelli:2002adven} we give the first determination of $f_{B_c}$ from
lattice QCD in the quenched approximation.

\section{Conclusions}
\label{sec:conclutions}

We have shown the effectiveness of the step scaling
method in the computation of the heavy--light and heavy--heavy
meson spectrum.
We have performed the calculations for the pseudoscalar and vector
mesons in the continuum limit of lattice QCD in the quenched
approximation and, by comparison with
the experimental determinations, we have extracted
the masses of the bottom and of the charm quarks.\\
Our results
represent the first determination of the bottom quark mass in the
continuum limit of quenched lattice QCD. 

Using the results on the quark masses we give an independent theoretical
calculation of the $B_c$ meson mass that matches very well the
experimental result.

\begin{ack}
We warmly thank A.~Kronfeld and R.~Sommer for fruitful discussions.
This work has been partially supported by the European Community 
under the grant HPRN--CT--2000--00145 Hadrons/Lattice QCD. 
\end{ack}

\bibliographystyle{h-elsevier} 
\bibliography{mcont}

\begin{table}[c]
\begin{center}
\scriptsize
\begin{tabular}{ccccc}
\midrule
$\qquad\qquad\beta\qquad\qquad$ & $\qquad m_1^{RGI}\qquad$ & $\qquad m_2^{RGI}\qquad$ & $\qquad M_P$ $\qquad$ &  $\qquad {M}_{V}$ $\qquad$ \\
\toprule
         &  7.14(8)             &  7.14(8)             &  8.295(5)          &  8.318(5)       \\
         &                      &  6.63(7)             &  8.092(5)          &  8.116(6)          \\
         &                      &  4.024(44)           &  6.898(6)          &  6.924(6)       \\
         &                      &  1.696(19)           &  5.590(7)          &  5.622(7)       \\
         &                      &  1.591(18)           &  5.527(7)          &  5.559(8)    \\
         &                      &  0.1381(30)          &  4.646(9)          &  4.691(10)   \\
         &                      &  0.0978(28)          &  4.622(10)         &  4.668(10)    \\
         &                      &  0.0574(28)          &  4.597(10)         &  4.645(10)     \\
\cmidrule{2-5}
         &  6.63(7)             &  6.63(7)             &  7.888(5)          &  7.913(6)       \\
         &                      &  4.024(44)           &  6.693(6)          &  6.721(6)       \\
         &                      &  1.696(19)           &  5.384(7)          &  5.417(7)          \\
         &                      &  1.591(18)           &  5.320(7)          &  5.354(8)       \\
         &                      &  0.1381(30)          &  4.438(9)          &  4.485(10)   \\
         &                      &  0.0978(28)          &  4.414(10)         &  4.462(10)   \\
         &                      &  0.0574(28)          &  4.390(10)         &  4.440(10)    \\
\cmidrule{2-5}
6.963    &  4.024(44)          &   4.024(44)           &  5.491(6)          &  5.524(7)       \\
         &                     &   1.696(19)           &  4.170(7)          &  4.215(8)          \\
         &                     &   1.591(18)           &  4.106(7)          &  4.152(8)   \\
         &                     &   0.1381(30)          &  3.207(10)         &  3.278(11)   \\
         &                     &   0.0978(28)          &  3.183(10)         &  3.255(11)    \\
         &                     &   0.0574(28)          &  3.158(10)         &  3.232(11)   \\
\cmidrule{2-5}
         &  1.696(19)          &   1.696(19)           &  2.826(8)          &  2.898(9)       \\
         &                     &   1.591(18)           &  2.760(8)          &  2.835(9)       \\
         &                     &   0.1381(30)          &  1.816(10)         &  1.952(12)   \\
         &                     &   0.0978(28)          &  1.790(10)         &  1.928(12)   \\
         &                     &   0.0574(28)          &  1.763(10)         &  1.905(12)    \\
\cmidrule{2-5}
         &  1.591(18)          &   1.591(18)           &  2.694(8)          &  2.771(9)       \\
         &                     &   0.1381(30)          &  1.746(10)         &  1.888(12)   \\
         &                     &   0.0978(28)          &  1.719(10)         &  1.864(12)   \\
         &                     &   0.0574(28)          &  1.692(10)         &  1.840(12)    \\
\bottomrule
\end{tabular}
\mycaption{Values of the pseudoscalar, $M_P$, and vector, $M_V$, meson masses resulting from the simulations on the smallest volume
 $L_0 = 0.4$ fm at $\beta=6.963$. Units are in GeV.}
\label{tab:SVNumRes}
\end{center}
\end{table}

\begin{table}
\begin{center}
\scriptsize
\begin{tabular}{ccccc}
\midrule
$\qquad\qquad\beta\qquad\qquad$ & $\qquad m_1^{RGI}\qquad$ & $\qquad m_2^{RGI}\qquad$ & $\qquad M_P$ $\qquad$ &  $\qquad {M}_{V}$ $\qquad$ \\
\toprule
         &  7.11(8)             & 7.11(8)        &  8.920(6)    &  8.944(7)       \\
         &                      & 6.61(20)       &  8.678(6)    &  8.702(7)       \\
         &                      & 4.018(44)      &  7.310(7)    &  7.336(8)       \\
         &                      & 1.695(19)      &  5.926(9)    &  5.957(10)       \\
         &                      & 1.592(18)      &  5.862(9)    &  5.893(10)       \\
         &                      & 0.1374(27)     &  4.987(13)   &  5.031(14)      \\
         &                      & 0.0971(24)     &  4.964(14)   &  5.009(14)      \\
         &                      & 0.0567(24)     &  4.942(14)   &  4.987(15)      \\
\cmidrule{2-5}
         &  6.61(20)            & 6.61(20)       &  8.435(6)    &  8.459(7)       \\
         &                      & 4.018(44)      &  7.066(7)    &  7.093(8)       \\
         &                      & 1.695(19)      &  5.680(9)    &  5.713(10)      \\
         &                      & 1.592(18)      &  5.616(9)    &  5.649(10)      \\
         &                      & 0.1374(27)     &  4.739(13)   &  4.786(14)   \\
         &                      & 0.0971(24)     &  4.717(14)   &  4.764(15)   \\
         &                      & 0.0567(24)     &  4.694(14)   &  4.742(15)    \\
\cmidrule{2-5}
7.300    &  4.018(44)           & 4.018(44)      &  5.689(8)    &  5.722(9)       \\
         &                      & 1.695(19)      &  4.291(10)   &  4.336(11)         \\
         &                      & 1.592(18)      &  4.226(10)   &  4.272(11)   \\
         &                      & 0.1374(27)     &  3.334(14)   &  3.404(15)   \\
         &                      & 0.0971(24)     &  3.310(14)   &  3.381(15)    \\
         &                      & 0.0567(24)     &  3.287(14)   &  3.359(16)   \\
\cmidrule{2-5}
         &  1.695(19)           & 1.695(19)      &  2.869(11)   &  2.942(13)       \\
         &                      & 1.592(18)      &  2.803(11)   &  2.878(13)       \\
         &                      & 0.1374(27)     &  1.864(14)   &  1.999(17)   \\
         &                      & 0.0971(24)     &  1.838(14)   &  1.976(17)   \\
         &                      & 0.0567(24)     &  1.812(15)   &  1.953(17)    \\
\cmidrule{2-5}
         &  1.592(18)           & 1.592(18)      &  2.736(11)   &  2.813(13)       \\
         &                      & 0.1374(27)     &  1.793(14)   &  1.934(17)   \\
         &                      & 0.0971(24)     &  1.767(14)   &  1.911(17)   \\
         &                      & 0.0567(24)     &  1.741(15)   &  1.888(18)    \\
\bottomrule
\end{tabular}
\mycaption{Values of the pseudoscalar, $M_P$, and vector, $M_V$, meson masses resulting from the simulations on the smallest volume
 $L_0 = 0.4$ fm at $\beta=7.300$. Units are in GeV.}
\label{tab:SVNumRes2}
\end{center}
\end{table}

\begin{table}
\begin{center}
\scriptsize
\begin{tabular}{ccccc}
\midrule
$\qquad\qquad\beta\qquad\qquad$ & $\qquad m_1^{RGI}\qquad$ & $\qquad m_2^{RGI}\qquad$ & $\qquad M_P$ $\qquad$ &  $\qquad {M}_{V}$ $\qquad$ \\
\toprule
         &  7.10(8)             & 7.10(8)            &  9.203(7)    &  9.225(8)          \\         
         &                      & 6.60(7)            &  8.939(7)    &  8.960(8)   \\                 
         &                      & 4.016(44)          &  7.480(8)    &  7.503(9)          \\         
         &                      & 1.698(19)          &  6.060(10)   &  6.088(10)   \\                 
         &                      & 1.595(18)          &  5.996(10)   &  6.025(11)   \\                 
         &                      & 0.1422(27)         &  5.115(14)   &  5.157(15)   \\                 
         &                      & 0.1021(25)         &  5.093(14)   &  5.135(15)   \\                 
         &                      & 0.0618(23)         &  5.070(15)   &  5.113(15)   \\                 
\cmidrule{2-5}			   	                                                                        
         &  6.60(7)             & 6.60(7)            &  8.674(7)    &  8.696(8)          \\         
         &                      & 4.016(44)          &  7.214(8)    &  7.238(9)          \\         
         &                      & 1.698(19)          &  5.792(10)   &  5.822(11)   \\                 
         &                      & 1.595(18)          &  5.728(10)   &  5.759(12)   \\                 
         &                      & 0.1422(27)         &  4.846(14)   &  4.891(15)   \\                 
         &                      & 0.1021(25)         &  4.823(14)   &  4.869(15)   \\                 
         &                      & 0.0618(23)         &  4.801(15)   &  4.847(15)   \\                 
\cmidrule{2-5}			   	                                                                        
7.548    &  4.016(44)           & 4.016(44)          &  5.746(9)    &  5.776(10)          \\         
         &                      & 1.698(19)          &  4.314(10)   &  4.356(11)   \\                 
         &                      & 1.595(18)          &  4.249(10)   &  4.292(12)   \\                 
         &                      & 0.1422(27)         &  3.352(15)   &  3.420(15)   \\                 
         &                      & 0.1021(25)         &  3.328(15)   &  3.398(16)   \\                 
         &                      & 0.0618(23)         &  3.305(15)   &  3.377(16)   \\                 
\cmidrule{2-5}			   	                                                                        
         &  1.698(19)           & 1.698(19)          &  2.860(11)   &  2.930(13)   \\                 
         &                      & 1.595(18)          &  2.794(11)   &  2.866(13)   \\                 
         &                      & 0.1422(27)         &  1.852(15)   &  1.988(17)    \\                 
         &                      & 0.1021(25)         &  1.827(15)   &  1.966(17)    \\                 
         &                      & 0.0618(23)         &  1.801(16)   &  1.944(18)   \\                 
\cmidrule{2-5}			   	                                                                        
         &  1.595(18)           & 1.595(18)          &  2.727(12)   &  2.802(13)   \\                 
         &                      & 0.1422(27)         &  1.781(15)   &  1.924(17)   \\                 
         &                      & 0.1021(25)         &  1.756(15)   &  1.901(18)   \\                 
         &                      & 0.0618(23)         &  1.730(16)   &  1.879(18)   \\                 

\bottomrule
\end{tabular}
\mycaption{Values of the pseudoscalar, $M_P$, and vector, $M_V$, meson masses resulting from the simulations on the smallest volume
 $L_0 = 0.4$ fm at $\beta=7.548$. Units are in GeV.}
\label{tab:SVNumRes3}
\end{center}
\end{table}

\begin{table}
\begin{center}
\scriptsize
\begin{tabular}{ccccc}
\midrule
$\qquad\qquad\beta\qquad\qquad$ & $\qquad m_1^{RGI}\qquad$ & $\qquad m_2^{RGI}\qquad$ & $\qquad \Sigma_P$ $\qquad$ &  $\qquad \Sigma_V$ $\qquad$ \\
\toprule
      & 0.1028(21)	  & 0.1028(21)	  	& 0.996(28)   & 0.826(21)\\
      & 	          & 0.2003(28)		& 1.010(23)   & 0.848(19)\\
      & 	          & 0.3007(36)		& 1.017(20)   & 0.869(17)\\
      & 		  & 1.927(21) 	  	& 1.034(7)    & 0.998(8)\\
      & 	          & 3.114(34) 		& 1.032(5)    & 1.013(6)\\
      & 		  & 3.543(39)  		& 1.031(5)    & 1.015(5)\\
\cmidrule{2-5}
      & 0.2003(28)        & 0.2003(28)		& 1.016(20)   & 0.867(17)\\
      &         	  & 0.3007(36)		& 1.020(18)   & 0.885(15)\\
      & 		  & 1.927(21) 	  	& 1.036(7)    & 1.003(7)\\
      & 	          & 3.114(34) 		& 1.034(5)    & 1.015(5)\\
      & 		  & 3.543(39)  		& 1.037(5)    & 1.018(5)\\
\cmidrule{2-5}
6.420 & 0.3007(36)	  & 0.3007(36)		& 1.022(16)   & 0.901(14)\\
      & 		  & 1.927(21) 	  	& 1.038(6)    & 1.007(7)\\
      & 	          & 3.114(34) 		& 1.036(5)    & 1.019(5)\\
      & 		  & 3.543(39) 		& 1.0349(46)  & 1.0205(47)\\
\cmidrule{2-5}
      & 1.927(21)         & 1.927(21)           & 1.0520(34)  & 1.0425(37)\\
      &                   & 3.114(34)           & 1.0493(27)  & 1.0438(29)\\
      &                   & 3.543(39)           & 1.0479(25)  & 1.0433(27)\\
\cmidrule{2-5}
      & 3.114(34)         & 3.114(34)          & 1.0471(21)  & 1.0440(23)\\
      &                   & 3.543(39)          & 1.0460(20)  & 1.0433(22)\\
\cmidrule{2-5}
      & 3.543(39)         & 3.543(39)          & 1.0449(19)  & 1.0427(20)\\
\bottomrule
\end{tabular}
\mycaption{Values of the step scaling functions for the pseudoscalar, $\Sigma_P$, 
and vector, $\Sigma_V$, meson masses resulting from the simulation at $\beta=6.420$ of the first evolution step $L_0\to L_1$. 
Units are in GeV.}
\label{tab:S1NumRes1}
\end{center}
\end{table}

\begin{table}
\begin{center}
\scriptsize
\begin{tabular}{ccccc}
\midrule
$\qquad\qquad\beta\qquad\qquad$ & $\qquad m_1^{RGI}\qquad$ & $\qquad m_2^{RGI}\qquad$ & $\qquad \Sigma_P$ $\qquad$ &  $\qquad \Sigma_V$ $\qquad$ \\
\toprule
      & 0.1080(21)	  & 0.1080(21)		& 0.907(25)   & 0.789(19)\\
      & 		  & 0.2090(28)	  	& 0.935(21)   & 0.810(17)\\
      & 	          & 0.3127(38)		& 0.951(19)   & 0.830(16)\\
      & 		  & 1.979(22)	  	& 1.009(7)    & 0.976(7)\\
      & 	          & 3.136(34)		& 1.014(5)    & 0.995(5)\\
      & 		  & 3.542(39) 		& 1.014(5)    & 0.998(5)\\
\cmidrule{2-5}
      & 0.2090(28)	  & 0.2090(28)	  	& 0.950(18)   & 0.828(16)\\
      & 		  & 0.3127(38)		& 0.960(16)   & 0.846(14)\\
      & 		  & 1.979(22)    	& 1.011(6)    & 0.980(7)\\
      & 	          & 3.136(34)   	& 1.0155(47)  & 0.998(5)\\
      & 		  & 3.542(39) 		& 1.0158(43)  & 1.0009(46)\\
\cmidrule{2-5}
6.737 & 0.3127(38)	  & 0.3127(38)		& 0.967(14)   & 0.862(13)\\
      & 		  & 1.979(22) 	  	& 1.014(6)    & 0.985(6)\\
      & 	          & 3.136(34) 		& 1.0176(44)  & 1.0013(47)\\
      & 		  & 3.542(39)     	& 1.0178(40)  & 1.0039(43)\\
\cmidrule{2-5}
      & 1.979(22)	  & 1.979(22) 	  	& 1.0376(30)  & 1.0279(33)\\
      & 	          & 3.136(34) 		& 1.0375(23)  & 1.0317(25)\\
      & 		  & 3.542(39) 		& 1.0367(21)  & 1.0317(23)\\
\cmidrule{2-5}
      & 3.136(34)         & 3.136(34)		& 1.0372(18)  & 1.0338(19)\\
      & 		  & 3.542(39)		& 1.0365(16)  & 1.0336(18)\\
\cmidrule{2-5}
      & 3.542(39)         & 3.542(39)		& 1.0358(15)  & 1.0333(17)\\
\bottomrule
\end{tabular}
\mycaption{Values of the step scaling functions for the pseudoscalar, $\Sigma_P$, 
and vector, $\Sigma_V$, meson masses resulting from the simulation at $\beta=6.737$ of the first evolution step $L_0\to L_1$. 
Units are in GeV.}
\label{tab:S1NumRes2}
\end{center}
\end{table}

\begin{table}
\begin{center}
\scriptsize
\begin{tabular}{ccccc}
\midrule
$\qquad\qquad\beta\qquad\qquad$ & $\qquad m_1^{RGI}\qquad$ & $\qquad m_2^{RGI}\qquad$ & $\qquad \Sigma_P$ $\qquad$ &  $\qquad \Sigma_V$ $\qquad$ \\
\toprule
      & 0.1086(20)	  & 0.1086(20)		& 0.914(28)   & 0.783(24)\\
      & 		  & 0.2112(28)	  	& 0.939(24)   & 0.807(21)\\
      & 	          & 0.3134(38)		& 0.953(21)   & 0.830(19)\\
      & 		  & 2.003(22)	  	& 1.012(8)    & 0.981(8)\\
      & 	          & 3.153(35)		& 1.016(6)    & 0.999(6)\\
      & 		  & 3.549(39) 		& 1.016(5)    & 1.002(6)\\
\cmidrule{2-5}
      & 0.2112(28)        & 0.2112(28)		& 0.951(21)   & 0.828(19)\\
      & 		  & 0.3134(38)		& 0.960(18)   & 0.847(17)\\
      & 		  & 2.003(22)    	& 1.013(7)    & 0.984(8)\\
      & 	          & 3.153(35)  		& 1.017(5)    & 1.001(6)\\
      & 		  & 3.549(39) 		& 1.017(5)    & 1.003(5)\\
\cmidrule{2-5}
6.963 & 0.3134(38)	  & 0.3134(38)		& 0.967(16)   & 0.864(16)\\
      & 		  & 2.003(22) 	  	& 1.015(7)    & 0.989(7)\\
      & 	          & 3.153(35) 		& 1.018(5)    & 1.004(5)\\
      & 		  & 3.549(39)  		& 1.0181(46)  & 1.006(5)\\
\cmidrule{2-5}
      & 2.003(22)         & 2.003(22)          & 1.0365(35)  & 1.0281(39)\\
      &                   & 3.153(35)          & 1.0357(27)  & 1.0309(30)\\
      &                   & 3.549(39)          & 1.0348(25)  & 1.0307(28)\\
\cmidrule{2-5}
      & 3.153(35)         & 3.153(35)          & 1.0349(21)  & 1.0321(23)\\
      &                   & 3.549(39)          & 1.0341(20)  & 1.0317(22)\\
\cmidrule{2-5}
      & 3.549(39)         & 3.549(39)          & 1.0333(19)  & 1.0313(20)\\
\bottomrule
\end{tabular}
\mycaption{Values of the step scaling functions for the pseudoscalar, $\Sigma_P$, 
and vector, $\Sigma_V$, meson masses resulting from the simulation at $\beta=6.963$ of the first evolution step $L_0\to L_1$. 
Units are in GeV.}
\label{tab:S1NumRes3}
\end{center}
\end{table}

\begin{table}
\begin{center}
\scriptsize
\begin{tabular}{ccccc}
\midrule
$\qquad\qquad\beta\qquad\qquad$ & $\qquad m_1^{RGI}\qquad$ & $\qquad m_2^{RGI}\qquad$ & $\qquad \Sigma_P$ $\qquad$ &  $\qquad \Sigma_V$ $\qquad$ \\
\toprule
      & 0.1466(44)	  & 0.1466(44)	  	 & 1.269(13)   & 1.308(15)\\
      & 	          & 0.2322(28)		 & 1.244(11)   & 1.286(12)\\
      & 	          & 0.3154(36)		 & 1.226(9)    & 1.267(10)\\
      & 		  & 1.337(15)	  	 & 1.1406(40)  & 1.1636(45)\\
      & 	          & 1.551(17)		 & 1.1332(37)  & 1.1539(42)\\
      & 		  & 2.012(22) 		 & 1.1214(32)  & 1.1383(36)\\
\cmidrule{2-5}
      & 0.2322(28)        & 0.2322(28)		 & 1.224(9)    & 1.266(10)\\
      &         	  & 0.3154(36)		 & 1.209(8)    & 1.249(9)\\
      & 		  & 1.337(15)  	  	 & 1.1337(35)  & 1.1557(40)\\
      & 	          & 1.551(17)  		 & 1.1268(32)  & 1.1467(37)\\
      & 		  & 2.012(22) 		 & 1.1158(29)  & 1.1322(32)\\
\cmidrule{2-5}
5.960 & 0.3154(36)	  & 0.3154(36)		 & 1.196(7)    & 1.235(8)\\
      & 		  & 1.337(15) 	  	 & 1.1278(32)  & 1.1489(37)\\
      & 	          & 1.551(17) 		 & 1.1213(30)  & 1.1405(34)\\
      & 		  & 2.012(22) 		 & 1.1110(27)  & 1.1269(30)\\
\cmidrule{2-5}
      & 1.337(15)         & 1.337(15) 	  	 & 1.0894(19)  & 1.029(21)\\
      & 	          & 1.551(17) 		 & 1.0853(18)  & 1.0978(20)\\
      & 		  & 2.012(22) 		 & 1.0785(16)  & 1.0895(17)\\
\cmidrule{2-5}
      & 1.551(17)         & 1.551(17)		 & 1.0814(17)  & 1.0931(18)\\
      & 		  & 2.012(22)		 & 1.0749(15)  & 1.0852(16)\\
\cmidrule{2-5}
      & 2.012(22)	  & 2.012(22)		 & 1.0689(13)  & 1.0782(14)\\
\bottomrule
\end{tabular}
\mycaption{Values of the step scaling functions for the pseudoscalar, $\Sigma_P$, 
and vector, $\Sigma_V$, meson masses resulting from the simulation at $\beta=5.960$ of the second evolution step $L_1\to L_2$. 
Units are in GeV.}
\label{tab:S2NumRes1}
\end{center}
\end{table}

\begin{table}
\begin{center}
\scriptsize
\begin{tabular}{ccccc}
\midrule
$\qquad\qquad\beta\qquad\qquad$ & $\qquad m_1^{RGI}\qquad$ & $\qquad m_2^{RGI}\qquad$ & $\qquad \Sigma_P$ $\qquad$ &  $\qquad \Sigma_V$ $\qquad$ \\
\toprule
      & 0.1510(19)	  & 0.1510(19)	  	 & 1.156(14)   & 1.255(18)\\
      & 	          & 0.2550(29)		 & 1.151(11)   & 1.235(15)\\
      & 	          & 0.3493(39)		 & 1.146(10)   & 1.219(12)\\
      & 		  & 1.389(15)   	 & 1.1035(46)  & 1.131(6)\\
      & 	          & 1.584(17)   	 & 1.1098(42)  & 1.123(5)\\
      & 		  & 1.984(22)  		 & 1.0899(37)  & 1.1095(44)\\
\cmidrule{2-5}
      & 0.2550(29)        & 0.2550(29)		 & 1.147(10)   & 1.218(12)\\
      &         	  & 0.3493(39)		 & 1.142(8)    & 1.204(10)\\
      & 		  & 1.389(15) 	  	 & 1.1010(40)  & 1.1258(48)\\
      & 	          & 1.584(17) 		 & 1.0961(37)  & 1.1181(44)\\
      & 		  & 1.984(22) 		 & 1.0879(33)  & 1.1056(39)\\
\cmidrule{2-5}
6.211 & 0.3493(39)	  & 0.3493(39)		 & 1.138(7)    & 1.193(9)\\
      & 		  & 1.389(15) 	  	 & 1.0983(37)  & 1.1210(44)\\
      & 	          & 1.584(17) 		 & 1.0936(34)  & 1.1137(40)\\
      & 		  & 1.984(22) 		 & 1.0857(30)  & 1.1020(35)\\
\cmidrule{2-5}
      & 1.389(15)	  & 1.389(15) 	  	 & 1.0727(20)  & 1.0842(24)\\
      & 	          & 1.584(17) 		 & 1.0695(19)  & 1.0799(22)\\
      & 		  & 1.984(22) 		 & 1.0640(17)  & 1.0728(20)\\
\cmidrule{2-5}
      & 1.584(17)         & 1.584(17)		 & 1.0664(18)  & 1.0759(21)\\
      & 		  & 1.984(22)		 & 1.0611(16)  & 1.0692(18)\\
\cmidrule{2-5}
      & 1.984(22)	  & 1.984(22)		 & 1.0564(14)  & 1.0633(16)\\
\bottomrule
\end{tabular}
\mycaption{Values of the step scaling functions for the pseudoscalar, $\Sigma_P$, 
and vector, $\Sigma_V$, meson masses resulting from the simulation at $\beta=6.211$ of the second evolution step $L_1\to L_2$. 
Units are in GeV.}
\label{tab:S2NumRes2}
\end{center}
\end{table}

\begin{table}
\begin{center}
\scriptsize
\begin{tabular}{ccccc}
\midrule
$\qquad\qquad\beta\qquad\qquad$ & $\qquad m_1^{RGI}\qquad$ & $\qquad m_2^{RGI}\qquad$ & $\qquad \Sigma_P$ $\qquad$ &  $\qquad \Sigma_V$ $\qquad$ \\
\toprule
      & 0.1055(15)	  & 0.1055(15)           & 1.167(27)   & 1.259(35)\\
      & 	          & 0.2038(24)	         & 1.162(22)   & 1.235(27)\\
      & 	          & 0.3016(34)	         & 1.156(18)   & 1.218(22)\\
      & 		  & 1.355(14)  	         & 1.104(7)    & 1.125(9)\\
      & 	          & 1.547(17)	         & 1.098(7)    & 1.117(8)\\
      & 		  & 1.933(21) 	         & 1.088(6)    & 1.103(7)\\
\cmidrule{2-5}
      & 0.2038(24)        & 0.2038(24)		 & 1.155(18)   & 1.217(21)\\
      &         	  & 0.3016(34)		 & 1.150(15)   & 1.203(18)\\
      & 		  & 1.355(14) 	  	 & 1.101(6)    & 1.121(8)\\
      & 	          & 1.547(17)   	 & 1.095(6)    & 1.113(7)\\
      & 		  & 1.933(21) 		 & 1.086(5)    & 1.000(6)\\
\cmidrule{2-5}
6.420 & 0.3016(34)	  & 0.3016(34)		 & 1.144(13)   & 1.192(16)\\
      & 		  & 1.355(14) 	  	 & 1.098(6)    & 1.118(7)\\
      & 	          & 1.547(17) 		 & 1.093(5)    & 1.110(6)\\
      & 		  & 1.933(21) 		 & 1.0839(46)  & 1.098(5)\\
\cmidrule{2-5}
      & 1.355(14)	  & 1.355(14) 	  	 & 1.0711(31)  & 1.0821(37)\\
      & 	          & 1.547(17) 		 & 1.0676(29)  & 1.0777(34)\\
      & 		  & 1.933(21) 		 & 1.0617(25)  & 1.0702(30)\\
\cmidrule{2-5}
      & 1.547(17)         & 1.547(17)		 & 1.0643(27)  & 1.0735(31)\\
      & 		  & 1.933(21)		 & 1.0587(24)  & 1.0665(27)\\
\cmidrule{2-5}
      & 1.933(21)	  & 1.933(21)		 & 1.0538(21)  & 1.0604(24)\\
\bottomrule
\end{tabular}
\mycaption{Values of the step scaling functions for the pseudoscalar, $\Sigma_P$, 
and vector, $\Sigma_V$, meson masses resulting from the simulation at $\beta=6.420$ of the second evolution step $L_1\to L_2$. 
Units are in GeV.}
\label{tab:S2NumRes3}
\end{center}
\end{table}

\end{document}